\documentclass{elsart}
\makeatletter
\def\proof{\@ifnextchar[{\@opargbeginproof}{\@beginproof}}
\def\@beginproof{\rm\parskip\z@\trivlist \item[\hskip \labelsep{\bf Proof}]}
\def\@opargbeginproof[#1]{\rm\parskip\z@\trivlist
      \item[\hskip \labelsep{\bf Proof\ #1}]}

\makeatother 
\usepackage{cite,graphicx,psfrag,amssymb,subfigure,url,verbatim}
\newtheorem{theorem}{Theorem}
\newtheorem{result}{Result}
\newtheorem{definition}{Definition}

\def\definedas{\triangleq}
\def\order{O}
\def\boldl{{\mbox{\boldmath $l$}}}
\def\boldsl{{\mbox{\scriptsize \boldmath $l$}}}
\def\Cost{L}

\def\w{w}
\def\q{a}

\def\ll{(}
\def\lr{)}

\def\Z{{\mathbb Z}}
\def\lg{{\log_2}}

\begin{document}
\begin{frontmatter}

\title{Alphabetic Coding with Exponential Costs\thanksref{titlefn}}
\thanks[titlefn]{Material in this paper was presented at the 2006
  International Symposium on Information Theory, Seattle, Washington,
  USA.}
\author{Michael~B.~Baer}
\ead{calbear@{\bf \tiny \.{1}}eee.org}

\begin{abstract}
An alphabetic binary tree formulation applies to problems in which an outcome needs to be determined via alphabetically ordered search prior to the termination of some window of opportunity.  Rather than finding a decision tree minimizing $\sum_{i=1}^n w(i) l(i)$, this variant involves minimizing $\log_\q \sum_{i=1}^n w(i) \q^{l(i)}$ for a given $\q \in (0,1)$.  This note introduces a dynamic programming algorithm that finds the optimal solution in polynomial time and space, and shows that methods traditionally used to improve the speed of  optimizations in related problems, such as the Hu-Tucker procedure, fail for this problem. This note thus also introduces two approximation algorithms which can find a suboptimal solution in linear time (for one) or $\order(n \log n)$ time (for the other), with associated coding redundancy bounds. 
\end{abstract}

\end{frontmatter}

\section{Introduction} 

Applications such as searching\cite{Knu71} and coding theory\cite{Huff}
make extensive use of binary trees.  We denote the length (number of edges) of a path
from the root to node~$i \in \{1, 2, \ldots, n\}$ of the tree as
$l(i)$, and the weight (usually probability) of the leaf as $w(i)$.  Given a
set of weights, Huffman's algorithm\cite{Huff} finds a tree minimizing cost
function 
\begin{equation}
\sum_{i=1}^n w(i)l(i)
\label{linear}
\end{equation}
and Hu and Tucker's algorithm \cite{HKT} finds an optimal
\textit{alphabetic} tree:
\begin{definition}
An \textrm{alphabetic} tree is a tree with leaves are in numerical
order given inorder tree traversal (i.e., $1,2,\ldots,n$ from left to
right).
\end{definition}

Three papers independently considered the problem of minimizing
\begin{equation}
\Cost_\q(\w,\boldl) \definedas \log_\q \sum_{i=1}^n w(i) \q^{l(i)}
\label{norm}
\end{equation}
for $\q>1$\cite[p.~254]{HKT} \cite[p.~485]{Park} \cite[p.~231]{Humb2}
for unconstrained (Huffman-like) minimization, the solution of which
is very similar to that of Huffman's algorithm.  One of these further
noted that an algorithm similar to Hu and Tucker's solves the
alphabetically constrained version of this problem \cite{HKT}, while another
noted that the Huffman-like solution also solves the unconstrained
(\ref{norm}) for $\q<1$ \cite{Humb2}, in which $\log_\q x$ is monotonically
decreasing and the objective's summation term is thus maximized.

A recent paper showed that the $\q<1$ problem describes certain
situations of single-shot decision-making\cite{Baer07}. Given a window
of time corresponding to a memoryless random variable, if we wish to
find the leaf of the binary tree through constant-time edge traversal,
this is found in time with probability $\q^{\Cost_\q(\w,\boldsl)}$ ---
which we thus wish to minimize --- for some known $\q<1$.  However,
solving the alphabetic version of this problem remained unaddressed.

Here we present an $\order(n^3)$ algorithm for minimizing (\ref{norm})
that is somewhat similar to Gilbert and Moore's method for
(\ref{linear})\cite{GiMo}.  We then introduce counterexamples on
attempts to minimize using faster methods, such the modification of Hu
and Tucker's, which only succeeds for $\q>1$.  Finally we present
approximation algorithms, related to those for the linear problem,
which find suboptimal solutions in $\order(n)$ and $\order(n \log n)$,
leading to simple bounds for both these solutions and the optimal
ones.

\section{Optimal Alphabetic Trees}
\label{alpha}

Because the alphabetic tree imposes leaf order, each decision of
which child to take, represented by a $0$ (for left) or $1$ (for
right), is equivalent to a question of the form, ``Is the output
greater than or equal to $s$?'' where $s$ is one of the possible
symbols, a symbol we call the \textit{splitting point}:

\begin{definition}
The \textrm{splitting point} of an internal node (or the corresponding
subtree) is the smallest index among the leaves of the right subtree.
\end{definition}
\begin{definition}
Each \textrm{codeword}~$c(i)$ is the sequence of bits corresponding to
the sequence of decisions (path) to arrive at leaf~$i$.  The overall
set of codewords --- alphabetic code~$C$ --- fully describes the tree,
as does length vector~$\boldl$, the sequence of lengths~$\{l(i)\}$.
\end{definition}

The dynamic programming approach of Gilbert and Moore\cite{GiMo} is
adapted to this problem (\ref{norm}):
\begin{theorem}
An algorithm
finds the maximum tree weight $W_{j,k}$ (and corresponding
optimum tree) for items $j$ through $k$ for each value of $k-j$ from
$0$ to $n-1$ (in order), by computing inductively 
\begin{equation}
\textstyle W_{j,k} \leftarrow \q \max_{s \in \{j+1,j+2,\ldots,k\}}
           [W_{j,s-1}+W_{s,k}] \mbox{ starting with } W_{j,j}
           \leftarrow w(j)
\label{dp}
\end{equation}
 for $1 \leq j<k \leq n$ in $\order(n^3)$ time and
$\order(n^2)$ space.
\end{theorem}

\begin{proof}
Recall first that maximizing $\q^{\Cost_\q(\w,\boldsl)} = \sum_i
w(i)\q^{l(i)}$ minimizes $\Cost_\q(\w,\boldl)$, which is why
(\ref{dp}) is a maximization operation.  One can see that $W_{1,n} =
\q^{\Cost_\q(\w,\boldsl)}$ inductively by considering a (sub)tree's
two subtrees as independent, rooted trees, one with summation
$W_{1,s-1} = \sum_{i=1}^{s-1} w(i)\q^{l(i)-1}$, the other with
summation $W_{s,n} = \sum_{i=s}^n w(i)\q^{l(i)-1}$.  Then $W_{1,n} =
\q (W_{1,s-1} + W_{s,n})$.  Starting with $W_{j,j} = w(j)$, then, we
see that these values can be built up accordingly (since the path
length from a leaf to itself is $0$, $W_{j,j} = \q^0w(j)$, and there
is nothing numerically special about the final tree).  Since all
subtrees of an optimal tree are optimal --- via a substitution
argument, e.g., \cite{Knu71} --- the maximization finds an optimal
solution.  This suggests the dynamic programming algorithm; similarly
to \cite{Knu71}, calculating all optimal subtrees of a size less than
that of the (sub)tree in a current step, we can try all possible
splitting points using optimal subtrees, yielding the optimal tree.

$\order(n^2)$ items are stored --- $\order(n^2)$ weights for every
possible range and the associated splitting points; these are used to
recursively find the implied subtree --- calculated by testing
$\order(n)$ splitting points for each internal node, thus
the time and space complexity.
\end{proof}

Knuth \cite{Knu71} reduced the algorithmic complexity of Gilbert and
Moore's method for (\ref{linear}) by using the fact that the splitting
point of an optimal tree of size $n$ must be between the splitting
points of the two optimal subtrees of size $n-1$.  With (\ref{norm}),
this no longer holds.  Consider $\q = 0.6$ with input weights $w=(8,
1, 9, 6)$.  The splitting point of $(8, 1, 9)$ is $s=3$
($w(s)=w(3)=9$, yielding subtrees with $(8,1)$ and $(9)$), and the
splitting point of $(1, 9, 6)$ is $s=4$ ($w(s)=6$).  However, the
optimal splitting point of $(8, 1, 9, 6)$ is $s=2$ ($w(s)=1$).

Similarly, for (\ref{norm}) with $\q > 1$\cite{HKT}, there is a
procedure based on the Hu-Tucker algorithm for finding an optimal
alphabetic solution. The Hu-Tucker algorithm begins with the input
weights arranged as leaves in numerical order ($1,2,\ldots,n$ in a
line).  It then combines the two items $i$ and $j$ that, of all pairs
of items without a leaf separating them, have a minimum weight sum,
putting it in the place of either node, both of which are now
(ordered) children.  In the original Hu-Tucker algorithm, this item is
given weight $w(i)+w(j)$, whereas for (\ref{norm}) with $\q > 1$ it is
given weight $\q w(i)+\q w(j)$.  The algorithm then finds the minimum
weighted pair among those pairs of distinct items (uncombined input
leaves and combined items) without any \textit{uncombined} leaf
between them, placing the resulting node in the place of either
original node.  Continuing on, we obtain a tree that is not
necessarily alphabetical, but which has the same lengths as an
alphabetic tree which can be easily reconstructed, (optimally) solving
the problem (for $\q>1$).

However, consider again $\q=0.6$, this time for weights $(8, 1, 9, 6,
2)$.  The Hu-Tucker-like algorithm first combines $6$ and $2$, then
$8$ and $1$, then the first combined node with $9$, and finally the
remaining two nodes, resulting in a tree with lengths $\boldl' = \ll
2, 2, 2, 3, 3 \lr$ and $\Cost_\q(\w,\boldl')\approx -4.121$.  However,
a tree with lengths $\boldl'' = \ll 1, 3, 3, 3, 3 \lr$, having
$\Cost_\q(\w,\boldl'') \approx -4.232$, shows that the Hu-Tucker-like
solution is nonoptimal.

\begin{result}
Knuth's method for speeding up dynamic programming fails for $\q<1$,
as does using the Hu-Tucker-like method optimal for $\q>1$.
\end{result}

\section{Approximation Algorithms and Bounds}
\label{appx}

In this section, we add the assertion $\sum_{i=1}^n w(i) = 1$ to our
problem, which can be considered an optimization of (\ref{norm}) with
constraints:
\begin{enumerate}
\item The Kraft inequality of binary trees, $\sum_{i=1}^n
2^{-l(i)}\leq 1$;
\item The integer constraint, $l(i) \in \Z$;
\item The alphabetic constraint.  
\end{enumerate}
The first and second of these are necessary and sufficient for the
lengths to correspond to a binary tree.  Relaxing the second and third
allows for a numerical solution which can bound the performance of the
optimal solution.  The numerical solution, $l^\dagger$, shown by
Campbell \cite{Camp0,Camp}, results in the Shannon-like
$$l^{\mbox{\scriptsize s}}(i) \definedas \lceil l^\dagger \rceil = \left\lceil
-\frac{1}{1+\lg \q} \lg w(i) + \lg \left(\sum_{j=1}^n
w(j)^{\frac{1}{1+\lg \q}}\right) \right\rceil$$ a valid (but not
necessarily optimal) solution to the problem with only the alphabetic
constraint relaxed, that is, the Huffman-like problem.

The approximation algorithm in Fig.~\ref{aalg} has a linear-time
variant patterned after that in \cite{Yeu1} --- relying on
$\boldl^{\mbox{\scriptsize s}}$ --- and a $\order(n \log n)$-time
variant patterned after \cite{Naka} --- instead using
$\boldl^{\mbox{\scriptsize h}}$, those lengths obtained from solving
the optimal code tree for the Huffman-like problem.

\begin{figure}
\begin{center}
{\bf Procedure for Finding a Near-Optimal Code}
\end{center}
\begin{enumerate}
\item Start with an optimal or near-optimal nonalphabetic code with
  length vector $\boldl^{\mbox{\scriptsize non}}$, either the
  Shannon-like $\boldl^{\mbox{\scriptsize s}}$ or the Huffman-like
  $\boldl^{\mbox{\scriptsize h}}$.
\item Find the set of all {\it minimal points}: $i$ such that $1<i<n$,
  $l^{\mbox{\scriptsize non}}(i)<l^{\mbox{\scriptsize non}}(i-1)$, and
  $l^{\mbox{\scriptsize non}}(i)<l^{\mbox{\scriptsize non}}(i+1)$; or
  $i \in [j, j+k]$ minimizing $w(i)$ for $l^{\mbox{\scriptsize
    non}}(j-1)>l^{\mbox{\scriptsize non}}(j)=l^{\mbox{\scriptsize
    non}}(j+1)=\cdots = l^{\mbox{\scriptsize non}}(j+k) <
  l^{\mbox{\scriptsize non}}(j+k+1)$.
\item Assign a preliminary alphabetic code with lengths
  $l^{\mbox{\scriptsize pre}}(i)=l^{\mbox{\scriptsize non}}(i)+1$ for
  all minimal points and $l^{\mbox{\scriptsize
      pre}}(i)=l^{\mbox{\scriptsize non}}(i)$ for all other items.
  The first codeword is $l^{\mbox{\scriptsize pre}}(1)$ zeros, and
  each additional codeword $c(i)$ is obtained by either truncating
  $c(i-1)$ to $l^{\mbox{\scriptsize pre}}(i)$ bits and adding $1$ to
  the integer that the binary codeword represents (if
  $l^{\mbox{\scriptsize pre}}(i)\leq l^{\mbox{\scriptsize pre}}(i-1)$)
  or by adding $1$ to the integer/codeword $c(i-1)$ and appending
  $l^{\mbox{\scriptsize pre}}(i)-l^{\mbox{\scriptsize pre}}(i-1)$
  zeros (if $l^{\mbox{\scriptsize pre}}(i) > l^{\mbox{\scriptsize
      pre}}(i-1)$), defining the binary tree.
\item Go through the code tree (with, e.g., a depth-first search), and
  remove any redundant nodes.  Any node with only one child can
  replace the child by its grandchild or grandchildren.  At the end of
  this process, an alphabetic code with $\sum_{i=1}^n 2^{-l(i)}=1$ is
  obtained.
\end{enumerate}
\line(1,0){390} 
\label{aalg}
\end{figure}

Every step after the first takes linear time with linear space, thus
the overall complexity of the algorithms.  Step~3 is the method by
which Nakatsu showed that any nonalphabetic code can be made into an
alphabetic code with similar lengths\cite{Naka}.  (The use of weights
as a tie breaker and the nonlinearity of the problem do not change the
validity of the algorithm.)  Step~4 is the method by which Yeung
showed that any alphabetic code can be made into another alphabetic
code with $\sum_{i=1}^n 2^{-l(i)}=1$ without lengthening any
codewords\cite{Yeu1}.  Thus this is a hybrid and extension of these
two approaches.

For $\w = (8/26, 1/26, 9/26, 6/26, 2/26)$ with $\q=0.6$, applying the
Shannon-like version of this algorithm, we find that
$\boldl^{{\mbox{\scriptsize s}}}= \ll 2, 13, 1, 4, 10\lr$, preliminary
codeword lengths are $\boldl_{{\mbox{\scriptsize
      s}}}^{\mbox{\scriptsize pre}} = \ll 2, 13, 2, 4, 10\lr$, and the
preliminary code tree is as follows:
$$C = \ll 00, 01{\mathit{00000000000}}, 10, 11{\mathit{0}}0, 11{\mathit{0}}1{\mathit{000000}} \lr
$$
The italicized bits are redundant, and therefore so are the
corresponding nodes in the code tree.  They are thus removed in
Step~4, which means the final tree has lengths $\ll 2, 2, 2, 3, 3\lr$.
The probability of success is $\q^{\Cost_\q(\w,\boldsl)} \approx
0.316$ (${\Cost_\q(\w,\boldl)} \approx 0.851$), close to the optimal
probability of about $0.334$ (${\Cost_\q(\w,\boldl^*)} \approx
0.843$).  Using the Huffman-like approximation algorithm yields
$\boldl^{\mbox{\scriptsize h}} = \ll 2, 4, 1, 3, 4\lr$, a preliminary
tree of lengths $\boldl_{\mbox{\scriptsize h}}^{\mbox{\scriptsize
    pre}} = \ll 2, 4, 2, 3, 4\lr$, and an output tree with lengths
$\ll 2, 2, 2, 3, 3\lr$, which are identical to the above.  The same
probability mass function with $\q=0.7$ yields an optimal tree in the
Huffman-like version.  For $a \in (0.5,1)$, coding bounds follow from
these approaches:
\begin{theorem}
Let 
\begin{itemize} 
\item $\Cost_\q^{\mathrm{{\bar{a}}}}(\w)$ be the minimized (\ref{norm}) for the alphabetic problem, 
\item $\Cost_\q^{\mathrm{{\tilde{h}}}}(\w)$
be that obtained using the $\boldl^{\mathrm{{h}}}$-based approximation algorithm,
\item $\Cost_\q^{\mathrm{{\tilde{s}}}}(\w)$ 
be that obtained using the $\boldl^{\mathrm{{s}}}$-based approximation algorithm,
\item $\Cost_\q^{\mathrm{{non}}}(\w)=\Cost_\q(\w,\boldl^{\mathrm{{non}}})$, $\Cost_\q^{\mathrm{{s}}}(\w)=\Cost_\q(\w,\boldl^{\mathrm{{s}}})$, and $\Cost_\q^{\mathrm{{h}}}(\w)=\Cost_\q(\w,\boldl^{\mathrm{{h}}})$ (using those $\boldl$ values from Fig.~\ref{aalg}).
\end{itemize}
Then
\begin{equation}
H_\alpha(\w)\leq\Cost_\q^{\mathrm{{h}}}(\w)\leq\Cost_\q^{\mathrm{{\bar{a}}}}(\w)\leq\Cost_\q^{\mathrm{{\tilde{h}}}}(\w)<1+\Cost_\q^{\mathrm{{h}}}(\w)<2+H_\alpha(\w)
\label{alphabound}
\end{equation}
\begin{equation}
H_\alpha(\w)\leq\Cost_\q^{\mathrm{{h}}}(\w)\leq\Cost_\q^{\mathrm{{\bar{a}}}}(\w)\leq\Cost_\q^{\mathrm{{\tilde{s}}}}(\w)<1+\Cost_\q^{\mathrm{{s}}}(\w)<2+H_\alpha(\w)
\label{slikebound}
\end{equation}
where $H_\alpha(\w)$ is the R\'{e}nyi entropy for $\alpha = (1+\lg \q)^{-1}$:
$$H_\alpha(\w) = \frac{1}{1-\alpha}\lg \sum_{i=1}^n w(i)^\alpha = (\log_a 2\q)\left(\lg \sum_{i=1}^n w(i)^{\frac{1}{1+\lg \q}}\right)$$ 
\end{theorem}

\begin{proof}
  This is a corollary of Campbell's Shannon-like bounds for $a > 0.5$
  --- $H_\alpha(\w)\leq\Cost_\q^{\mbox{\scriptsize h}}(\w)\leq
  \Cost_\q^{\mbox{\scriptsize s}}(\w)<1+H_\alpha(\w)$ --- along with
  the facts that (a) the two approximation algorithm lengths
  corresponding to items $1$ and $n$ are no greater than those in
  $\boldl^{\mbox{\scriptsize non}}$ and (b) no other length exceeds
  the corresponding length in $\boldl^{\mbox{\scriptsize non}}$ by $1$
  or more.  This results in $\Cost_\q^{\mbox{\scriptsize
      \~{h}}}(\w)<1+\Cost_\q^{\mbox{\scriptsize h}}(\w)$ and
  $\Cost_\q^{\mbox{\scriptsize
      \~{s}}}(\w)<1+\Cost_\q^{\mbox{\scriptsize s}}(\w)$ due to
  (\ref{norm}), and, since no alphabetic tree is better than the
  optimal alphabetic tree and no alphabetic tree is better than the
  optimal Huffman-like tree, we arrive at (\ref{alphabound}) and
  (\ref{slikebound}).
\end{proof}
The lower limit to $\Cost_\q^{\mathrm{{\bar{a}}}}(\w)$ is satisfied by
$(2, 2)$, while the upper limit is approached by $(\epsilon,
1-2\epsilon, \epsilon)$, which approaches entropy $0$ and penalty~$2$.

Both these algorithms and the bounds due to analogous inequalities
apply to $\q>1$ and to the traditional alphabetic problem ($\q
\rightarrow 1$, where $H_1$ is Shannon entropy \cite{Camp}).  For the
traditional problem, due to Step~4, the Huffman-based approximation
version of the above algorithm is a strict improvement on Yeung's
Huffman-based approximation \cite{Yeu1}.

\section*{Acknowledgments}
The author would like to thank T.~C.~Hu and J.~David Morgenthaler for
discussions and encouragement on this topic.

\ifx \cyr \undefined \let \cyr = \relax \fi


\begin{thebibliography}{10}

\bibitem{Baer07}
M.~B. Baer.
\newblock Optimal prefix codes for infinite alphabets with nonlinear costs.
\newblock {\em IEEE Trans. Inf. Theory}, IT-54(3):1273--1286, Mar. 2008.

\bibitem{Camp0}
L.~L. Campbell.
\newblock A coding problem and {R{\'{e}}nyi's} entropy.
\newblock {\em Inf. Contr.}, 8(4):423--429, Aug. 1965.

\bibitem{Camp}
L.~L. Campbell.
\newblock Definition of entropy by means of a coding problem.
\newblock {\em Z. Wahrscheinlichkeitstheorie und verwandte Gebiete},
  6:113--118, 1966.

\bibitem{GiMo}
E.~N. Gilbert and E.~F. Moore.
\newblock Variable-length binary encodings.
\newblock {\em Bell Syst. Tech. J.}, 38:933--967, July 1959.

\bibitem{HKT}
T.~C. Hu, D.~J. Kleitman, and J.~K. Tamaki.
\newblock Binary trees optimum under various criteria.
\newblock {\em SIAM J. Appl. Math.}, 37(2):246--256, Apr. 1979.

\bibitem{Huff}
D.~A. Huffman.
\newblock A method for the construction of minimum-redundancy codes.
\newblock {\em Proc. IRE}, 40(9):1098--1101, Sept. 1952.

\bibitem{Humb2}
P.~A. Humblet.
\newblock Generalization of {Huffman} coding to minimize the probability of
  buffer overflow.
\newblock {\em IEEE Trans. Inf. Theory}, IT-27(2):230--232, Mar. 1981.

\bibitem{Knu71}
D.~E. Knuth.
\newblock Optimum binary search trees.
\newblock {\em Acta Informatica}, 1:14--25, 1971.

\bibitem{Naka}
N.~Nakatsu.
\newblock Bounds on the redundancy of binary alphabetical codes.
\newblock {\em IEEE Trans. Inf. Theory}, IT-37(4):1225--1229, July 1991.

\bibitem{Park}
D.~S. Parker, Jr.
\newblock Conditions for optimality of the {Huffman} algorithm.
\newblock {\em SIAM J. Comput.}, 9(3):470--489, Aug. 1980.

\bibitem{Yeu1}
R.~W. Yeung.
\newblock Alphabetic codes revisited.
\newblock {\em IEEE Trans. Inf. Theory}, IT-37(3):564--572, May 1991.

\end{thebibliography}
\end{document}